\documentclass{ws-mplb}
\usepackage{mathrsfs}
\begin{document}

\markboth{Armin Rahmani}{Quantum evolution with an ensemble of Hamiltonians}

\catchline{}{}{}{}{}
%
%%%%%%%%%%%%%%%%%%%%%%%%%%%%%%%%%%%%%%%%%%%%%%%%%%%%%%%%%%%%%%%%%%%%

\title{QUANTUM DYNAMICS WITH AN ENSEMBLE OF HAMILTONIANS
}

\author{\footnotesize Armin Rahmani\footnote{}}

\address{Theoretical Division, T-4 and CNLS, Los Alamos National Laboratory\\ Los Alamos, New Mexico 87545, USA\\
armin@lanl.gov}

\maketitle

\begin{history}
\received{(Day Month Year)}
\revised{(Day Month Year)}
\end{history}

\begin{abstract}
We review recent progress in the nonequilibrium dynamics of thermally isolated many-body quantum systems, evolving with an \textit{ensemble} of Hamiltonians as opposed to deterministic evolution with a single time-dependent Hamiltonian. Such questions arise in (i) quantum dynamics of disordered systems, where different realizations of disorder give rise to an ensemble of real-time quantum evolutions. (ii) quantum evolution with noisy Hamiltonians (temporal disorder), which leads to stochastic Schr\"odinger equations, and, (iii) in the broader context of quantum optimal control, where one needs to  analyze an ensemble of permissible protocols in order to find one that optimizes a given figure of merit. The theme of ensemble quantum evolution appears in several emerging new directions in noneqilibrium quantum dynamics of thermally isolated many-body systems, which include many-body localization, noise-driven systems, and shortcuts to adiabaticity.

\end{abstract}

\keywords{nonequilibrium dynamics; optimal control; many-body localization; noise.}

\section{Introduction}

Nonequilibrium many-body quantum dynamics of thermally isolated systems has come to the forefront of research in atomic, molecular, optical and condensed matter physics in recent years.\cite{Greiner2002,Tuchman2006,Kinoshita2006,Sadler2006,Hofferberth2007} The remarkable progress in cooling, trapping and manipulating atoms has made it possible to create closed quantum systems with custom-made Hamiltonians and long coherence times.\cite{Lewenstein2007,Bloch2008} These developments have provided a fertile playground to experimentally study real-time quantum dynamics of many-body systems, which has in turn sparked great theoretical interest.\cite{Polkovnikov2011}

Unlike equilibrium physics, where a general framework for studying phases of quantum matter [in terms of renormalization group (RG) and universality classes] has been developed, real-time dynamics is still a field in its infancy. This provides an exciting opportunity for posing novel questions, searching for new universal behavior, and ultimately devising a unified framework for understanding nonequilibrium quantum matter. The focus of zero-temperature equilibrium many-body quantum physics is the ground state (which may also encode some information about the low-energy excited states) of an interacting many-body Hamiltonian. The properties of the ground-state wave function (e.g., order parameters, correlation functions, entanglement characteristics) determine a phase diagram as a function of parameters in the Hamiltonian. Thus, the primary challenge of equilibrium many-body physics is solving (or determining some properties of the solution of) the time-independent Schr\"odnger equation. In nonequilibrium dynamics, such Hamiltonian parameters are functions of time (rather than single numbers), and the time-dependent  Schr\"odnger equation determines the final state of the system as a function of the initial state and the trajectory by which these parameters change.

The large body of work in recent years indicates that a plethora of interesting phenomena occur in closed many-body quantum systems with time-dependent Hamiltonian parameters even if the time dependence is as simple as a sudden quench of a single coupling constant.\cite{Polkovnikov2011}
Most of the recent of work in quantum dynamics concerns deterministic evolution with a single prescribed protocol. At the core of these studies lies unitary evolution with a given Hamiltonian $H(t)$ and its effect on the many-body quantum wave function. We do not review such studies here as excellent reviews are already available.\cite{Polkovnikov2011} Instead, we focus on one emerging new direction where deterministic quantum evolution with a single Hamiltonian is not the whole story.

The simplest situation, in which we need to study an \textit{ensemble} of evolutions, is the real-time dynamics in (spatially) disordered systems. For example, one can envision two cases: (i) $H(t)=H_{\rm clean}(t)+H_{\rm disorder}$, (ii) $H(t)=H_{\rm clean}+H_{\rm disorder}(t)$. In case (i), the part of the Hamiltonian which changes with time is not disordered. Nevertheless, an ensemble of quantum evolutions is generated by the sample-to-sample variations of the time-independent part of the Hamiltonian. Case (i) is indeed closest in spirit to studying deterministic quantum evolution. However, we still need to develop techniques for disorder averaging of time-evolved observables. Such problems have provided useful probes\cite{Znidaric2008,Chiara2006,Bardarson2012,Khatami2012,Vosk2013,Serbyn2013,Huse2013} for the many-body localization transition\cite{Basko2006,Basko2007,Oganesyan2007,Pal2010}. Case (ii), on the other hand,  still remains largely unexplored to the best of our knowledge.  In case (ii), we do not change a single parameter (such as a uniform interaction strength) in a system that has spatial disorder in some other parameter, but, instead, make the disorder itself time dependent. As an example, one can imagine a scenario where each realization of spatial disorder is turned on linearly with time.

A second situation, in which ensemble quantum dynamics comes into play, involves disorder in the time domain, where some Hamiltonian parameter(s) change in time in a stochastic manner.\cite{Dalessio2013,Pichler2012,Pichler2013,DallaTorre2010,DallaTorre2012,Marino2012} As noise is inevitably present in many systems, e.g., the lasers forming an optical lattice in cold-atom experiments, such problems are of great practical and fundamental interest. Generically, a noisy Hamiltonian is expected to cause heating and change the behavior of the correlation functions and other observables. As the time dependence of the Hamiltonian has a stochastic nature, averaging over noise, i.e., trajectories of the time-dependent parameter, is the key new ingredient, which must be dealt with in the analysis of such systems.

Finally, there is a rather different set of problems, known as many-body quantum optimal control problems\cite{Doria2011,Rahmani2011}  (quantum optimal control has a long and rich history in few-body physics and in particular in controlling molecules\cite{Montangero2007,Brif2010,Peirce1988,Calarco2004,Hohenester2007,Grond2009}), in which we need to analyze (at least in some approaches) an ensemble of quantum evolutions. Even though we may not have any disorder, either in time or in space, we are in fact dealing with an ensemble of quantum evolutions as we are  interested in reverse engineering the problem: instead of asking what happens after the system evolves with a given prescribed protocol, we want to design/find a protocol that performs a specific task, i.e., creates a quantum state with certain desired characteristics (generally by optimizing a given figure of merit). A direct approach to such problems involves an unbiased search through all permissible protocols,\cite{Rahmani2011,Rahmani2013} and requires understanding the quantum dynamics of the ensemble of such protocols. There are many different approaches to designing quantum protocols for various purposes. In case of transforming ground states into one another, such problems are generally referred to as shortcuts to adiabaticity\cite{Torrontegui2013}. We do not discuss schemes other than optimal control in this article, which include invariant-based methods\cite{Chen2010}, fast-forward approach\cite{Masuda2010}, and transitionless quantum driving with nonlocal Hamiltonians.\cite{Berry2009}

In this article, we focus on the three directions outlined above. The review is not intended to be an exhaustive survey of the literature, but rather an overview of some of the recent ideas and techniques in ensemble quantum evolution. In Sec.~\ref{sec:disorder}, we briefly review recent work on real-time evolution of disordered systems as a probe of the many-body localization transition. Sec.~\ref{sec:noise} is devoted to the dynamics of stochastically driven systems. We discuss techniques such as the Fokker-Planck and the master-equation approach, and outline some recent results and developments. In Sec.~\ref{sec:optimal}, we turn to the problem of optimal control in quantum evolution, and, in particular, we discuss the method of simulated annealing for selecting optimal protocols out of an ensemble of permissible ones. We close the paper in Sec.~\ref{sec:conc} with a brief summary.

\section{Spatial disorder: time evolution as a probe of many-body localization}\label{sec:disorder}

It has been suggested that disorder can lead to localization of a closed many-body quantum system. Inhibition of energy flow prevents thermalization\cite{Rigol2008} even when the system has finite energy density.\cite{Anderson1958} This phenomenon is generally referred to as many-body localization.\cite{Basko2006}

The intimate relation between the flow of energy and many-body localization suggests that real-time dynamics may serve as a powerful probe of this phenomenon.\cite{Khatami2012} Several recent works have considered the growth of entanglement entropy in interacting disordered systems after a quantum quench from an unentangled initial state  (direct product of wave functions of local degrees of freedom).\cite{Znidaric2008,Chiara2006,Bardarson2012,Vosk2013,Serbyn2013,Huse2013} Growth of entanglement entropy after quantum quenches has been widely studied in clean systems.\cite{Calabrese2005,Fagotti2008,Sengupta2009,Rahmani2010} In system with ballistic energy transport, which do not exhibit many-body localization, the general picture for the growth of entanglement entropy after a sudden quantum quench from such a product state is as follows:\cite{Calabrese2005} the excess energy deposited into the system (after the quantum quench, the initial state is generally a superposition of ground and several excited states of the new Hamiltonian with some excess energy with respect to the ground state) flows through the system in quasiparticle excitations, which have a characteristic velocity. Therefore, for shorter times, the number of quasiparticles that reach subsystem $A$ (assumed to be the smaller subsystem) form subsystem $B$ is proportional to the time $t$. As pairs of quasiparticles emanating form a point in space are entangled, the generated entanglement entropy should also grow linearly with time, and eventually saturate to values proportional to the volume of the smaller subsystem. The volume scaling can be understood by noting that the growth should stop when quasiparticles from $B$ have reached every point in the volume of subsystem $A$. Another way to understand the volume scaling is through the assumption that the subsystem effectively experiences a heat bath from the rest of the system and eventually thermalizes to a reduced density matrix that is close a thermal density matrix, which has an extensive entropy. As discussed below, this picture breaks down for many-body localized systems. Interestingly, it was found recently, that even with diffusive energy transport (as opposed to ballistic), entanglement entropy can grow linearly with time.\cite{Kim2013} 
Note that the ballistically moving quasiparticles, which carry the entanglement as discussed above, appear in integrable systems. More generic systems thermalize through the diffusive dynamics of energy as the quasiparticles experience strong scattering.\cite{Kim2013}  Finally there are the many-body localized states, which do not thermalize.

It turns out that a global quantum quench in the presence of many-body localization also leads to the volume scaling of the saturation value of the entanglement entropy (albeit with a smaller coefficient), but the growth is generically \textit{logarithmic} in time instead of linear. This result was first found numerically in several one-dimensional systems\cite{Znidaric2008,Chiara2006,Bardarson2012} and a few recent works shed more light on the origin of this phenomenon.\cite{Vosk2013,Serbyn2013,Huse2013}  Before proceeding to explain the reasons behind the logarithmic growth, we mention an alternative static probe:\cite{Bauer2013} In the regime of many-body localization, most excited eigenstates respect the area law instead of the volume scaling. In clean systems, only the ground state and possibly some low-energy excited states respect the area law and most eigenstates have volume scaling. This statement is related to the so-called eigenstate thermalization hypothesis:\cite{Deutsch1991,Srednicki1994} an eigenstate with a finite energy density is very similar to a thermal state at a temperature corresponding to its energy density in the sense that it produces similar expectation values for local observables. This in turn implies that the the reduced density matrices of small enough subsystems resemble thermal density matrices $e^{-\beta H}/Z$, where $H$ is the Hamiltonian and $Z={\rm tr}\left(e^{-\beta H}\right)$ is the partition function. Such thermal density matrices have extensive entropy, which implies volume scaling of the entanglement entropy.

Some insight can be gained into the logarithmic growth of the entanglement entropy from a phenomenological approach\cite{Huse2013} (or a more quantitative version\cite{Serbyn2013} not discussed here). In the localized phase, the eigenstates of the many-body Hamiltonian are expected to be direct products of certain exponentially localized degrees of freedom (not identical to the strictly local bare degrees of freedom, which would give rise to a vanishing entanglement entropy instead of an area law). Generically, such local degrees of freedom can be written as combinations of the bare degrees of freedom, say, spins, with  weights that decay exponentially with distance. Let us represent these local degrees of freedom by dressed spins $\tau^z$. As the Hamiltonian should commute with all $\tau_z$ (due to the fact that many-body eigenstates are constructed directly in terms of these dressed spins), we expect a phenomenological Hamiltonian\cite{Huse2013}
\begin{equation}
H=\sum_i h_i\tau^z_i+\sum_{ij}J_{ij}\tau^z_i\tau^z_j+\sum_{ijk}J_{ijk}\tau^z_i\tau^z_j\tau^z_k+\dots,
\end{equation}
where the coupling constants should generically fall off exponentially (they come from inserting the expression for the bare spins in terms of $\tau$ spins into a Hamiltonian that is strictly short-range, say, nearest-neighbor, in bare spins).
After a transient time, the unentangled initial state develops area-law entanglement. At later times, we expect from the Hamiltonian above that the dressed spins should precess around the $z$ axis with a rate set by the interactions with the other dressed spins. Let us focus on one-dimensional systems for simplicity. A spin $\tau_i$ becomes entangles with another spin $\tau_j$ after a time of order $t\propto 1/J_{ij}\propto e^{\ell_{ij}/\xi}$, where $\ell_{ij}$ is the distance between dressed spins $i$ and $j$, and $\xi$ is a characteristic correlation length. To reach an entanglement entropy $S$, we need to entangle dressed spins over a distance $\ell\propto S$, which leads to $t\propto e^{S/\xi}$, indicating generic logarithmic entanglement growth.

A useful technique for analyzing this problem quantitatively is a strong-disorder real-space renormalization group scheme, which was first developed for determining the ground-state properties of disordered spin chains\cite{Dasgupta1980,Bhatt1982,Fisher1994} (including to their entanglement entropy\cite{Refael2004}). The idea of such real-space RG, in the static case, is the successive decimation of the strongest bond in the system. In each decimation step, one approximates the system by its strongest coupling term and treats the rest of the chain as a perturbation. Writing out the change in the ground-state energy due to this perturbation leads to an effective interaction between the sites adjacent to the strongest bond. In other words, each RG step corresponds to removing sites $i$ and $i+1$ and connecting sites $i-1$ and $i+2$ by a renormalized effective coupling constant. These sites  are in turn relabeled as two consecutive sites. Repeated application of this RG decimation step changes the distribution of the coupling constants from their initial distribution. If a stationary distribution is reached eventually, we have found an RG fixed point.

Interestingly, such scheme can also be applied to the time evolution.\cite{Vosk2013} Once again, let us consider the bond with the strongest coupling constant $J_i$ for a Hamiltonian of type $\sum_i J_i\left(S^x_iS^x_{i+1}+S^y_iS^y_{i+1}+\Delta_iS^z_i S^z_{i+1}\right)$. For simplicity, consider an antiferromagnetic initial state. If we neglect all other terms in the Hamiltonian, and only keep the bond with the largest $|J_i|$, we have a $4\times 4$ Hamiltonian for the two spins coupled by this strongest bond. The initial state of these two spins is $|\uparrow\downarrow\rangle$, which is only connected by a zonzero Hamiltonian matrix element to itself and  $|\downarrow\uparrow\rangle$. Thus, the time evolution is simply given by a rotation with a characteristic frequency $\Omega \propto J_{\rm max}$ between these two states. For time scales larger than $\Omega^{-1}$, the fast oscillations average out the off-diagonal elements of the density matrix in the energy basis $|\pm\rangle=\left(|\uparrow\downarrow\rangle\pm |\downarrow\uparrow\rangle \right)/\sqrt{2}$, leaving behind the diagonal elements. The RG step then corresponds to such averaging over fast oscillations, removing the strongest bond and treating the couplings from the strongest bond to the rest of the chain as perturbation. More precisely, an effective Hamiltonian is constructed in each step that replaces the strongest bond with a bond connecting the two sites adjacent to it with renormalized coupling constants so as to reproduce the same time-evolved density matrix after averaging over fast oscillations. As expected, (even though the philosophy of this dynamical case is very different), the renormalized coupling constants bare a striking similarity to the classic static case, which targets the ground state.

An interesting novel feature of the dynamical case is that the effective Hamiltonian after the RG decimation depends on the starting state $|\pm\rangle$ of the strongest bond. The fact that the starting antiferromagnetic state is a superposition of $|\pm\rangle$ gives rise to the emergence of entanglement between the adjacent spins after the characteristic time scale of the decimation. It was shown,\cite{Vosk2013} through such explicit RG calculations, that in the generic case the entanglement entropy grows logarithmically with time. Other nongeneric behaviors, e.g., in the noninteracting case were also predicted from real-space RG.\cite{Vosk2013} For example, in case of random Ising spin chain, growth of the form $\left(\log t\right)^\alpha$, where $\alpha\neq1$ is a universal number, has been predicted at the critical point.\cite{Vosk2013b}

\section{Temporal disorder: time evolution with a noisy Hamiltonian}\label{sec:noise}

\subsection{General framework}

In this section, we consider the case where a parameter in the Hamiltonian fluctuates in time around some trajectory. For concreteness, we focus on a global parameter fluctuating around a constant value. (The more general case of fluctuations around a nontrivial protocol is of interest, e.g., in studying the robustness of optimal-control protocols\cite{Rahmani2011,Choi2012}) The general setup can then be written as 
\begin{equation} \label{eq:noise}
H(t)=H[g(t)], \qquad g(t)=g_0+\delta g(t),
\end{equation}
where $\delta (g)$ represents the noise, or, in other words, the temporal disorder of the Hamiltonian protocols. For concreteness, let us consider Gaussian white noise with $\overline {\delta g(t)}=0$ and
\begin{equation}  \label{eq:corr}
\overline{ \delta g(t)\delta g(t')}=W^2 \delta(t-t'),
\end{equation}
with $W$ representing the strength of noise and $\delta(t-t')$ the Dirac $\delta$ function. Assuming that the noise is small, the above scenario leads to a stochastic Schr\"odinger equation:
\begin{equation} \label{eq:sto_schro}
i\partial_t |\psi(t)\rangle=\left[H(g_0)+H'(g_0)\delta g(t)\right]|\psi(t)\rangle,
\end{equation}
where $H'(g)$ is shorthand for $\partial_g H(g)$. In addition to stemming from natural sources of fluctuations, such stochastic Hamiltonians can also be created by engineering random pulses as in recent experiments on superconducting qubits.\cite{Li2013} A general effect of such noisy driving is heating the system, i.e., increasing its energy content. One area where universal behavior may emerge is the fluctuations of this excess energy.\cite{Bunin2011}

We note that the fluctuations stem from two distinct sources:\cite{Dalessio2013}
\begin{romanlist}[(ii)]
\item An inherent quantum source: the final wave functions are quantum superpositions of energy eigenstates.
\item A classical source: different trajectories $\delta g(t)$ (realizations of temporal disorder) lead to realization-to-realization variations of the final wave functions.
\end{romanlist}
Determining the relative importance of these sources is of fundamental interest. As energy fluctuations are typically characterized by moments of energy, we can construct two distinct types of moments, which can help distinguish the effects of the sources above.\cite{Molmer1993,Dalessio2013}

The first type of energy moment we introduce encodes the realization-to-realization fluctuations of the quantum expectation values of energy in the final wave functions:
\begin{equation} \label{eq:Var1}
{\rm Var}_1(\epsilon)=\overline{\langle \epsilon\rangle^2}-\left(\overline{\langle \epsilon\rangle}\right)^2,
\end{equation}
where overlines (brackets) represent noise (quantum) averaging, and the expectation values are computed with respect to the constant Hamiltonian $H(g_0)$, i.e., for a given wave function $|\psi(t)\rangle$, $\langle \epsilon\rangle\equiv \langle \psi(t)|H(g_0)|\psi(t)\rangle$.  As the above expression characterizes the variations of a quantity, which is already quantum averaged, over different realizations of noise, the source (i) does not contribute to it. Therefore, it only captures the effect of source (ii). Experimentally, the measurement of the variance~(\ref{eq:Var1}) requires the repeated generation of identical realizations of noise (so for each realization a quantum average $\langle \epsilon\rangle$ can be determined). Such procedure may be implemented only for engineered noise as opposed to noise from natural sources. Thus, the above variation is not generically accessible in experiment. It serves, however, as a powerful theoretical diagnostic (which can be computed by analytical and numerical methods) for distinguishing the effects of sources (i) and (ii) above.

The second type of energy moment we introduce is an experimentally relevant one, which can be extracted from a histogram of the outcomes of energy measurements after evolution with different realizations of noise:\cite{Dalessio2013}
\begin{equation} \label{eq:Var2}
{\rm Var}_2(\epsilon)=\overline{\langle \epsilon^2\rangle}-\left(\overline{\langle \epsilon\rangle}\right)^2,
\end{equation}
where  for a given wave function $|\psi(t)\rangle$, $\langle \epsilon^2\rangle\equiv \langle \psi(t)|H^2(g_0)|\psi(t)\rangle$. As expected, the variance~(\ref{eq:Var2}) mixes the contributions of both (i) and (ii) sources above. Thus, the difference between Eqs.~(\ref{eq:Var1}) and ~(\ref{eq:Var2}) stems only from inherent quantum fluctuations.

\subsection{Fokker-Planck approach: application to a noisy Luttinger liquid}
\label{sec:FP}

For any system where quantum dynamics (with an arbitrary protocol) can be studied efficiently (either through exact solutions or numerical methods such as time-dependent density-matrix renormalization group), one can directly generate a large number of protocols, and perform the noise averaging after computing the wave function and/or the quantities of interest for each randomly generated protocol. To implement the noise in Eqs.~(\ref{eq:noise}) and ~(\ref{eq:corr}) in the numerics, one can discretize time and generate, e.g., piecewise constant protocols with $N$ pieces for time $0<t<\tau$ such that the constant $g(t)$ for each piece of duration $\Delta t=\tau/N$ is drawn from uniform distribution $[-{{\cal W}\over 2} , {{\cal W}\over 2} ]$, with ${\cal W}=\sqrt{12} {W \over \sqrt{\Delta t}}$. The results of such simulations are reliable if they are independent of $N$ so we need to keep increasing $N$ until the results converge.

While the numerical simulation of the stochastic Schr\"odinger equation provides a useful generic method, some rare exact results have been obtained for relatively simple models. Here we discuss two approaches based on the Fokker-Planck (FP) equation (the present section) and the master equation (next section) to this problem.
The idea of the FP approach is as follows. The wave function $|\psi(t)\rangle$ can always be written in terms of some parameters. For example, it can be expanded in a Hilbert space basis with an exponential number of amplitudes. In solvable models, we can have anz\"atse with far fewer number of parameters. Now by inserting the parameterization of the wave function into the stochastic Schr\"odinger equation, we obtain (generally coupled) stochastic differential equations (hereafter referred to as Langevin equations) for these parameters. Then, using the standard methods of classical nonequilibrium statistical mechanics, we can transform the set of Langevin equations into a FP equation, which governs the evolution of the joint probability distribution of the parameters determining the wave function. This probability distribution entails the probability of all possible wave functions, which can arise from the noisy time evolution. Before proceeding, let us restate the steps:\cite{Dalessio2013}
\begin{romanlist}[(ii)]
\item Writing the nonequilibrium wave function in terms of some parameters.
\item Inserting the parameterization of the wave function into the stochastic Schr\"odinger equation.
\item Transforming the Langevin equations into a FP equation for the wave-function probability distribution.
\end{romanlist}

To demonstrate the application of the FP approach, we focus on a particular model.\cite{Dalessio2013} A convenient system for studying the above questions analytically is a noise-driven Luttinger liquid with the following Hamiltonian:
\begin{equation}
H(K)=u\sum_{q>0}\left(K\:\Pi_{q}\Pi_{-q}+\frac{1}{K}\: q^{2}\:\Phi_{q}\Phi_{-q}\right),\label{eq:LL}
\end{equation}
where the Luttinger parameter $K$ encodes the strength of interactions, the velocity $u$ sets an overall energy scale (hereafter, we set $u$ to unity), and $\Pi_{q}$ is the conjugate momentum to bosonic field $\Phi_{q}$. The Hamiltonian above can be written as a sum of harmonic-oscillator Hamiltonians by breaking the field $\Phi_{q}$ into its real and imaginary components. Moreover, we can shift the Hamiltonian by a constant so the energies are measured with respect to the ground state.

We now consider a fluctuating Luttinger parameter $K(t)=K_0+\delta K(t)$. Experimentally, the Luttinger liquid may be realized, e.g., by an elongated quasicondensate with the interaction strength (and consequently the Luttinger parameter $K$) set by the transverse confinement. The fluctuations of lasers creating the confinement potential can lead to the appearance of fluctuations $\delta K(t)$. 
Assuming that the fluctuating lasers have spatial correlations longer than the length of the quasicondensate, the time-dependent Luttinger parameter is a global (spatially uniform) function of time.
For Hamiltonian~(\ref{eq:LL}), the following well-known fact from elementary quantum mechanics leads to a convenient parametrization of the nonequilibrium wave function: a Gaussian wave function evolving with an arbitrary time-dependent quadratic Hamiltonian retains it Gaussian form.\cite{Polkovnikov2008,Rahmani2011,Dalessio2013} Assuming that the system is initially in the ground state of $H(K_0)$ [see Eq.~(\ref{eq:LL})], the wave function for each momentum mode is initially a simple Gaussian. The time-dependent many-body wave function for an arbitrary protocol $K(t)$ can then be written as 
\begin{equation}
\Psi(\{\Phi_{q}\},t)=\prod_{q>0}\left(\frac{2\: q{\rm Re}\: z_{q}(t)}{\pi}\right)^{\frac{1}{2}}\exp\left[-q\: z_{q}(t)\:|\Phi_{q}|^{2}\right],\label{eq:psi}
\end{equation}
where $z_{q}(0)=K_{0}^{-1}$ [corresponding to the ground state of $H(K_0)$], and the functions $z_{q}(t)$ satisfy the Riccati equation
\begin{equation}\label{eq:ricc}
i\dot{z}_{q}(t)=\frac{q}{K(t)}\left\{[K(t)\, z_{q}(t)]^{2}-1\right\}.
\end{equation}

It is convenient to formulate the problem in terms of $z_q$ above and the fluctuations $\delta \alpha(t)$ of $\alpha(t)=1/K(t)$. We consider white noise for $\delta \alpha$ with strength $W$, i.e., substitute $g$ with $\alpha$ in Eq.~(\ref{eq:corr}). The Langevin equation for the dynamical variables $z_q$ can then be written as 
\begin{eqnarray}\label{eq:lang}
 \dot{\mathscr{R}}_{q}&=&2K_{0}q\:\mathscr{R}_{q}\mathscr{I}_{i}-2K_{0}^{2}q\:\mathscr{R}_{q}\mathscr{I}_{q}\:\delta\alpha,\\
\dot{\mathscr{I}}_{q}&=&K_{0}q\:\left(\mathscr{I}_{q}^{2}-\mathscr{R}_{q}^{2}+K_{0}^{-2}\right)-K_{0}^{2}q\:\left(\mathscr{I}_{q}^{2}-\mathscr{R}_{q}^{2}-K_{0}^{-2}\right)\:\delta\alpha,
\end{eqnarray}
where $\mathscr{R}_{q}\equiv {\rm Re}z_q$ and $\mathscr{I}_{q}\equiv {\rm Im}z_q$. We have broken up $z_q$ into real and imaginary parts in order to obtain real Langevin equations. 
After writing Langevin equations for the wave-function parameters, we need to write observables of interest (expectation value of energy or its higher moments, correlation functions, etc.) in terms of these parameters. The standard FP techniques of nonequilibrium classical statistical mechanics can now be used to compute the noise average of the observables of interest.

Before discussing the results for the Luttinger liquid, let us briefly review the classical FP approach. Consider a vector of dynamical variables $\vec{x}$ satisfying the Langevin equation
$\partial_{t}x_{i}=h_{i}(\vec{x})+g_{i}(\vec{x})\xi(t)$, with $\overline{\xi(t)\xi(t^{\prime})}=2\delta(t-t^{\prime})$. The probability distribution of these dynamical variables then evolves according to the FP equation~\cite{Risken1989}
\begin{equation}\label{eq:FP}
\partial_t f(\vec{x},t)=\left(-\frac{\partial}{\partial x_{i}}h_{i}-\frac{\partial}{\partial x_{i}}\frac{\partial g_{i}}{\partial x_{j}}g_{j}+\frac{\partial}{\partial x_{i}}\frac{\partial}{\partial x_{j}}g_{i}g_{j}\right)f(\vec{x},t),
\end{equation}
where summation over repeated indices is implied. To compute the noise-averaged expectation value of a function ${\cal F}(\vec{x})$ of the dynamical variables at time $t$, we can use the formal solution of the differential equation above [obtained by exponentiating the differential operator acting on the probability distribution on the right-hand side of Eq.~(\ref{eq:FP})], and repeated integrations by parts to write
\begin{equation}\label{eq:noise_ave}
\left.\overline{{\cal F}(\vec{x})}\right|_{t}=\int\prod_{i}\: d x_{i}\: f(\vec{x},0)\: \exp\left[\left(h_{i}\frac{\partial}{\partial x_{i}}+\frac{\partial g_{i}}{\partial x_{j}}g_{j}\frac{\partial}{\partial x_{i}}+g_{i}g_{j}\frac{\partial}{\partial x_{i}}\frac{\partial}{\partial x_{j}}\right)t\right]{\cal F}(\vec{x}).\label{eq:formal2}
\end{equation}

If at $t=0$ the system is in a particular state for all realizations of noise, the initial distribution function $f(\vec{x},0)$ is a delta function (the dynamical variables $x$ are the parameters in the wave function). Therefore, the integration is trivial once we compute the function obtained by acting with the exponential differential operator in Eq.~(\ref{eq:formal2}) on ${\cal F}(\vec{x})$. Generically, this is a complicated problem but a short-time expansion in $t$ can always be obtained straightforwardly by expanding the exponential operator in Eq.~(\ref{eq:noise_ave}). Interestingly, in case of the average energy,
\begin{equation}
{\cal E} ( \{ \mathscr{R}_{q} \},\{\mathscr{I}_{q}\})=\sum_q\frac{q}{2}\left[\frac{1}{2K_{0}\mathscr{R}_{q}}\left(1+K_{0}^{2}\left(\mathscr{R}_{q}^{2}+\mathscr{I}_{q}^{2}\right)\right)-1\right],
\end{equation}
of the Luttinger liquid evolving with the Langevin equation~(\ref{eq:lang}), the series obtained by this expansion can be resummed, which leads to the following exact result:~\cite{Dalessio2013}
\begin{equation}\label{eq:e_bar}
\overline{\langle\epsilon\rangle}=\frac{L}{8\pi K_{0}^{2}W^{2}t}\left(e^{2K_{0}^{2}\pi^{2}W^{2}t}-2\pi^{2}K_{0}^{2}W^{2}t-1\right),
\end{equation}
where the initial distribution function is $\delta(\mathscr{R}_{q}-K_0^{-1})\delta(\mathscr{I}_{q})$ from the corresponding initial conditions on $z_q$.We note that the expression above is obtained under the assumption that the Luttinger-liquid Hamiltonian is the correct description of the system at all energies. This assumption breaks down at high energies for realistic systems (the Luttinger liquid is only an effective low-energy Hamiltonian). In fact, an underlying assumption for th validity of Eq.~(\ref{eq:e_bar}) is the presence of a frequency cutoff of the order of the bandwidth in the noise spectrum. At shorter times, Eq.~(\ref{eq:e_bar}) exhibits linear growth of energy with a heating rate that readily follows from Eq.~(\ref{eq:e_bar}).

Since the experimentally interesting regime corresponds to small absorbed energy, we can study the fluctuations of energy using a perturbative approach based on expanding the Langevin equation in $\delta z_q=z_q-K_0^{-1}$, which leads to a simple linear Langevin equation
\begin{equation}
i\:\delta\dot{z}_{q}=2q\:\left(\delta z_{q}-\delta\alpha\right).
\end{equation}
In this limit, we can write an explicit integral expression for $\delta z_q(t)$ in terms of $\delta\alpha(t)$. The expressions for ${\rm Var}_1(\epsilon)$ and ${\rm Var}_2(\epsilon)$ can be written in terms of $z_q$ and expanded in $\delta z_q$. Due to the Gaussian nature of the noise, the noise averages of multi-point functions of $\delta \alpha$ break into sums of products of two-point functions (which are Dirac $\delta$ functions). The calculations are tedious but straightforward so we only quote the following results:\cite{Dalessio2013}
\begin{romanlist}[(ii)]
\item In the thermodynamic limit $L\rightarrow \infty$, ${\rm Var}_2(\epsilon)\approx {\rm Var}_2(\epsilon)$. In other words, \textit{the classical source originating from realization-to-realization variations of the final wave function dominates the energy fluctuations.}
\item Unlike thermodynamic equilibrium, where the relative fluctuations of energy $\sqrt{{\rm Var}(\epsilon)}/{\rm E}(\epsilon)\sim 1/\sqrt{L}$ [${\rm E}(\epsilon)$ is the average energy], in this noise-driven nonequilibrium system, \textit{the relative fluctuations do not decay with system size, but rather with time: $\sqrt{{\rm Var}(\epsilon)}/{\rm E}(\epsilon)\sim 1/\sqrt{t}$.}
\end{romanlist}

\subsection{Master-equation approach: application to noisy Bose-Hubbard model}
An alternative approach to thermally isolated noisy systems is the master-equation approach. 
The general formalism in the notation of Eqs.~(\ref{eq:corr}) and (\ref{eq:sto_schro}) states that $\overline{\rho (t)}=\overline{|\psi(t)\rangle\langle \psi(t)|}$ satisfies the following master equation:
\begin{equation}
\partial_t \overline{\rho(t)}=-i[H(g_0), \overline{\rho(t)}]-{W^2\over 2}[H'(g_0),[H'(g_0),\overline{\rho(t)}]].
\end{equation}
Let us give an elementary (not rigorous but intuitive) derivation of the above expression. The discretized approximation of white noise introduced in the beginning of Sec.~\ref{sec:FP} is very useful for this derivation. Let us introduce discrete stochastic variables $w_i$ with $\overline{w_i^2}=W^2$ such that the $\delta g(t)$ takes on the value $w_n/\sqrt{\Delta t}$ in the $n$th segment of the piecewise constant protocol. We can then write for times $t$ that are integer multiples of $\Delta t$:
\begin{equation}
\overline{\rho(t)}=\overline{\prod_{n=1}^N e^{iH(g_0)\Delta t+iw_nH'(g_0)\sqrt{\Delta t}}\times\rho(0)\times\prod_{m=1}^N e^{-iH(g_0)\Delta t-iw_mH'(g_0)\sqrt{\Delta t}}},
\end{equation}
where the limit $\Delta t=t/N\rightarrow 0$ will be taken at the end.
As different $w_n$ are uncorrelated, we can perform the noise averaging step by step and write
\begin{equation}
\overline{\rho(t)}=\overline{ e^{iH(g_0)\Delta t+iw_1H'(g_0)\sqrt{\Delta t}}\times\overline{\rho(t-\Delta t)}\times e^{-iH(g_0)\Delta t-iw_1H'(g_0)\sqrt{\Delta t}}}.
\end{equation}
The averaging above is now over a single stochastic variable $w_1$. Since we are interested in the limit of $\Delta t\rightarrow 0$, we can expand each evolution operator as
\begin{equation*}
e^{-iH(g_0)\Delta t-iw_1H'(g_0)\sqrt{\Delta t}}\approx 1-iH(g_0)\Delta t-iw_1H'(g_0)\sqrt{\Delta t}-{1\over 2}w_1^2H'^2(g_0)\Delta t+{\cal O}(\Delta t^{3/2}).
\end{equation*}
It is then easy to observe that to order $\Delta t$, three terms contribute to $\overline{\rho(t)}-\overline{\rho(t-\Delta t)}$, which are proportional to $w_1^2$ (note that $\overline {w_1}=0$). The averaging can then be done easily leading to the master equation.

Solving the master  equation above  (This is not generally straightforward and we may need to resort to direct simulation of the Langevin equation\cite{Pichler2013} in most cases) yields the noise-averaged density matrix at time $t$, which can in turn give the noise-averaged expectation values of different operators through
\begin{equation}
\overline {\langle O(t)\rangle}={\rm tr}\left[ \overline{\rho (t)} O \right].
\end{equation}
Notice that quantities like ${\rm Var}_1$ in Eq.~(\ref{eq:Var1}) or entanglement entropy can not be written in terms of the noise-averaged density matrix, which makes it necessary to use methods other than the master equation (such as the Fokker-Planck method of the previous section or direct Langevin simulations).

With a combination of analytical approximations and numerical methods such as time-dependent density-matrix renormalization group, the heating effects in noisy optical lattices have been studied recently with the master equation and direct Langevin simulations.\cite{Pichler2012,Pichler2013} The relevant model is the one-band Bose-Hubbard model
\begin{equation}
H(J, U)=-J\sum_{\langle ij\rangle}b^\dagger_ib_j+{U\over 2}\sum_i b^\dagger_ib^\dagger_i b_i b_i.
\end{equation}
Fluctuations in the laser forming the optical lattice lead to fluctuations in both the hopping amplitude $J$ and the interaction strength $U$, which to the first approximation are both functions of the potential depth $V$ of the wells in the optical lattice, $J=J(V)$ and $U=U(V)$.

 As in Eq.~(\ref{eq:sto_schro}), this scenario leads to the following stochastic Schr\"odinger equation:
\begin{align}\label{eqSMBSE}
\frac{d}{dt}|\Psi\rangle &  =-i\left[ H(J_{0},U_{0})+H\left(  \frac{dJ}%
{dV},\frac{dU}{dV}\right)  \delta V(t)\right] |\Psi\rangle, 
\end{align}
for $J_0=J(V_0)$, $U_0=U(V_0)$, and $V(t)=V_0+\delta V(t)$, with $\delta V(t)$ a Gaussian white noise. It was noted that if the additional term proportional to $\delta V$ commutes with the bare Hamiltonian $H(J_0, U_0)$, the noise just represents an overall Hamiltonian rescaling and does not lead to any excitations if the system is initially in the ground state of $H(J_0, U_0)$.\cite{Pichler2012} Such \textit{sweet spot} occurs if $U/J$ remains invariant with respect to the variations of $V$. An interesting proposal has been made for engineering such sweet spots in practice.\cite{Pichler2012} Additionally the heating rate has been computed recently in cases where we are away from the sweet spot.\cite{Pichler2013}

\section{Optimal control of quantum evolution: selecting optimal protocols out of a permissible ensemble}\label{sec:optimal}
\subsection{General approach}
\begin{figure}[th]
\centerline{\psfig{file=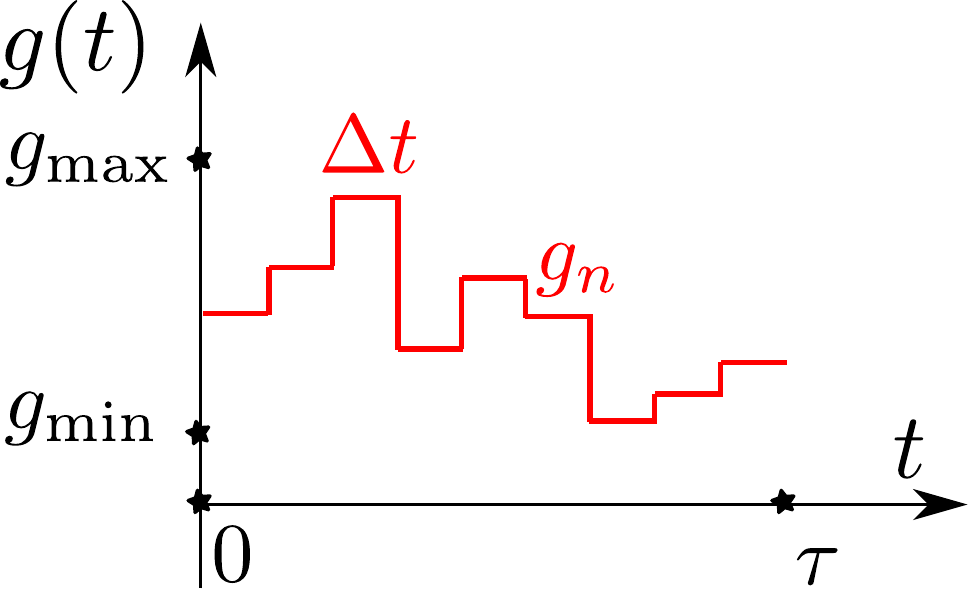,width=4.5cm}}
\vspace*{8pt}
\caption{Approximating an arbitrary permissible protocol with a piecewise constant protocol. In the limit of $\Delta t\rightarrow \infty$, we get an unbiased sampling. In the simulated annealing calculations, we need to increase the number of discrete intervals to reach convergence in the shape of the protocol obtained by the simulations. \label{f1}}
\end{figure}
Another area, where it is important to analyze an ensemble of Hamiltonians, is designing protocols that perform specific tasks. This is in some sense a question of reverse engineering; instead of asking what time evolution with a particular time-dependent Hamiltonian does, we are interested in finding/designing a protocol that produces a state with certain desired characteristics. One such example is the widely used adiabatic scheme, where one changes a Hamiltonian with the goal of taking the system from an eigenstate (say ground state) of an initial Hamiltonian to the ground state of the final Hamiltonian. According the the adiabatic theorem, if this process is done slowly enough (the time scale of the process is long compared with the inverse of the minimum energy gap to excitations), the adiabatic process can successfully transform the state with minimal excitations.

Of course, the adiabatic path is not always convenient as the required times can get too large (maybe even larger than the life time of the system). The adiabatic algorithm keeps the state on the ground state manifold during the evolution, which is an unnecessary restriction as we are only interested in the final state and not the transient states during the evolution. Optimal control, on the other hand, relaxes this restriction.

 The idea of optimal control is as follows.\cite{Rahmani2011} Suppose we have an ensemble of permissible protocols, which we can apply over a fixed time $\tau$ (set by the coherence time of the system). A very common ensemble corresponds to having upper and lower bounds on the value of the tunable Hamiltonian parameter $g(t)$: $g_{\rm  min}<g(t)<g_{\rm max}$, without any restriction on continuity or differentiability (we can have sudden quenches). Given an initial state (either a pure-state wave function or a mixed-state density matrix), each permissible protocol evolves the system deterministically resulting in a unique final wave function (or density matrix if we start with a mixed state). Let us assume our goal is to minimize a certain function of the final state.

If we can solve the time evolution for all permissible protocols, it has been shown recently that a simple numerical method known as simulated annealing is very effective in finding the optimal protocol out of all  permissible ones.\cite{Rahmani2011,Rahmani2013}  The method starts by approximating an arbitrary permissible protocol with a piecewise constant one over discrete time intervals as shown in Fig.~\ref{f1}. For a given finite $\Delta t$, this provides a biased optimization. However, an unbiased sampling can be obtained if we keep increasing the number of discrete intervals so that the shape of the obtained optimal protocol converges.\cite{Rahmani2011,Rahmani2013} 

For completeness, we describe this simple Monte-Carlo method in detail. We start from an arbitrary (say constant everywhere) protocol and compute the cost function ${\cal E}_1(\tau)$, i.e., the function we seek to minimize by designing an optimal protocol. We then select a random time interval $n$ and move the corresponding discrete control parameter $g_n$ up or down by a random small (compared with the range $g_{\rm max}-g_{\rm min}$) amount, computing the new cost function ${\cal E}_2(\tau)$ for the modified protocol. A simple Metropolis algorithm is then used: if ${\cal E}_2(\tau)<{\cal E}_1(\tau)$, we accept the move and keep the updated protocol, otherwise we accept the move with probability proportional to $e^{-\beta\left[{\cal E}_2(\tau)-{\cal E}_1(\tau)\right]}$, where $\beta$ is a fictitious inverse temperature. We then keep repeating the procedure until there is convergence in the shape of the protocol. As expected, the fictitious temperature $\beta$ should be varied as a function of the Monte-Carlo step $P$. Initially we want a small $\beta$ (compared with the characteristic $\left[{\cal E}_2(\tau)-{\cal E}_1(\tau)\right]^{-1}$ (at those initial stages of the Monte-Carlo simulations) so that most moves are initially accepted. Eventually, the fictitious temperature is reduced resulting in very large $\beta$ so only moves that improve the protocol are accepted. The dependence $\beta(P)$ is known as the annealing schedule. The efficiency and the success rate of the simulations depend on the judicious choice of annealing schedule. To find a good annealing schedule, we experimented with different exponents for a power-law dependence of $\beta(P)$ on $P$, and selected an exponent that produced the fastest convergence.\cite{Rahmani2011,Rahmani2013} 

Before discussing some of the results obtained with this method (e.g., a powerful scheme for extending the limits of atomic cooling), let us propose an approach, which my help extend the power of optimal control to generic experimental situations. As discussed above, the Monte-Carlo algorithm is extremely simple. The difficulty stems from solving the real-time quantum evolution and obtaining the cost function ${\cal E}(\tau)$ for a given (almost arbitrary) piecewise constant protocol. For simpler systems, one may be able to find exact or approximate (but controlled and reliable) analytical/numerical solutions  (for one-dimensional systems, efficient numerical methods such as time-dependent density-matrix-renormalization method may be combined with the simulated annealing method). However, the ultimate power of optimal control lies in its application to complex systems, which are not amenable to the existing analytical and numerical methods (take the two-dimensional fermionic Hubbard model as an example). Cold atoms provide a promising candidate for \textit{quantum simulations},\cite{Cirac2012} which may help us find solutions to these complex problems. In fact, there has been remarkable progress in creating model Hamiltonians corresponding to such complex systems. However, preparing interesting states (e.g., strongly correlated or highly frustrated quantum ground states) remains a challenge, partly due to the limitations of the existing cooling methods.

A key observation is that by integrating a classical computer, which performs the Monte-Carlo simulations, into the experiment, the system itself can be used to find its desired state. In other words, access to the system allows us to repeatedly initialize it in some state, evolve it with a given protocol, \textit{measure} the cost function ${\cal E}(\tau)$, and then feed this measured cost function into the computer running the Monte-Carlo simulation. This integration may also help with potential robustness issues stemming from inaccuracies in the generation of the protocol or the calibration of the instrument. This scheme is illustrated in Fig.~\ref{f2} below.
\begin{figure}[th]
\centerline{\psfig{file=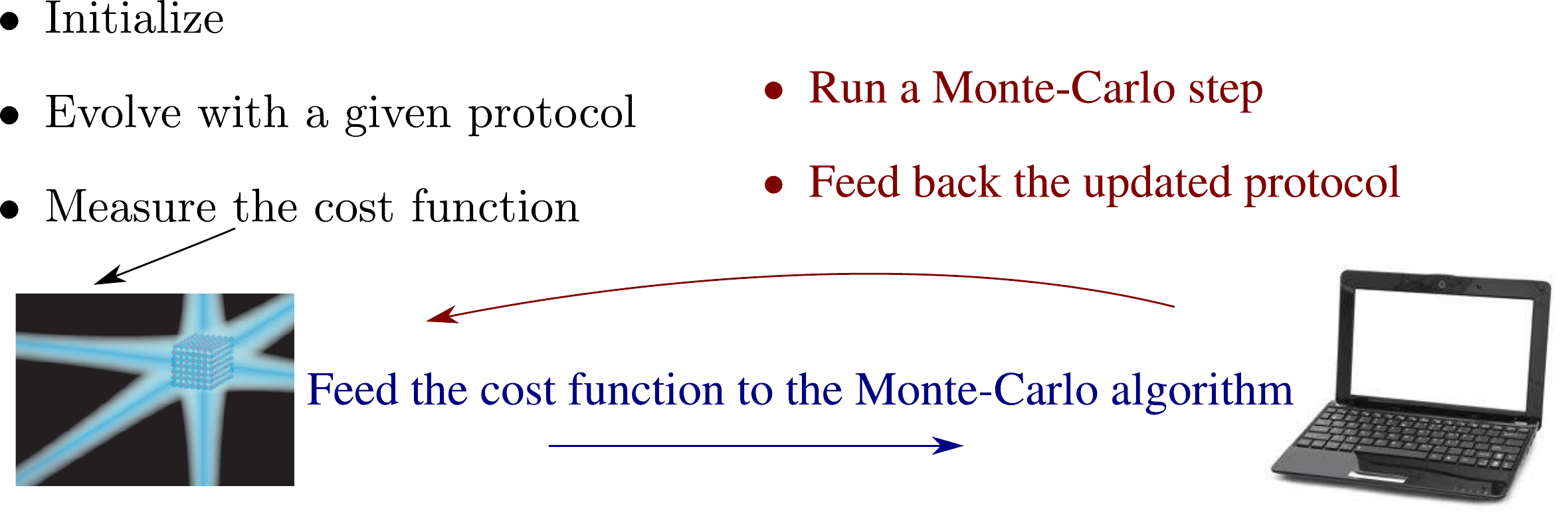,width=10cm}}
\vspace*{8pt}
\caption{Integrating the experiment, which provides the cost functions of trial protocols through evolution and measurement, with the Monte-Carlo simulations. \label{f2}}
\end{figure}

\subsection{Optimal control for state preparation}
\begin{figure}[th]
\centerline{\psfig{file=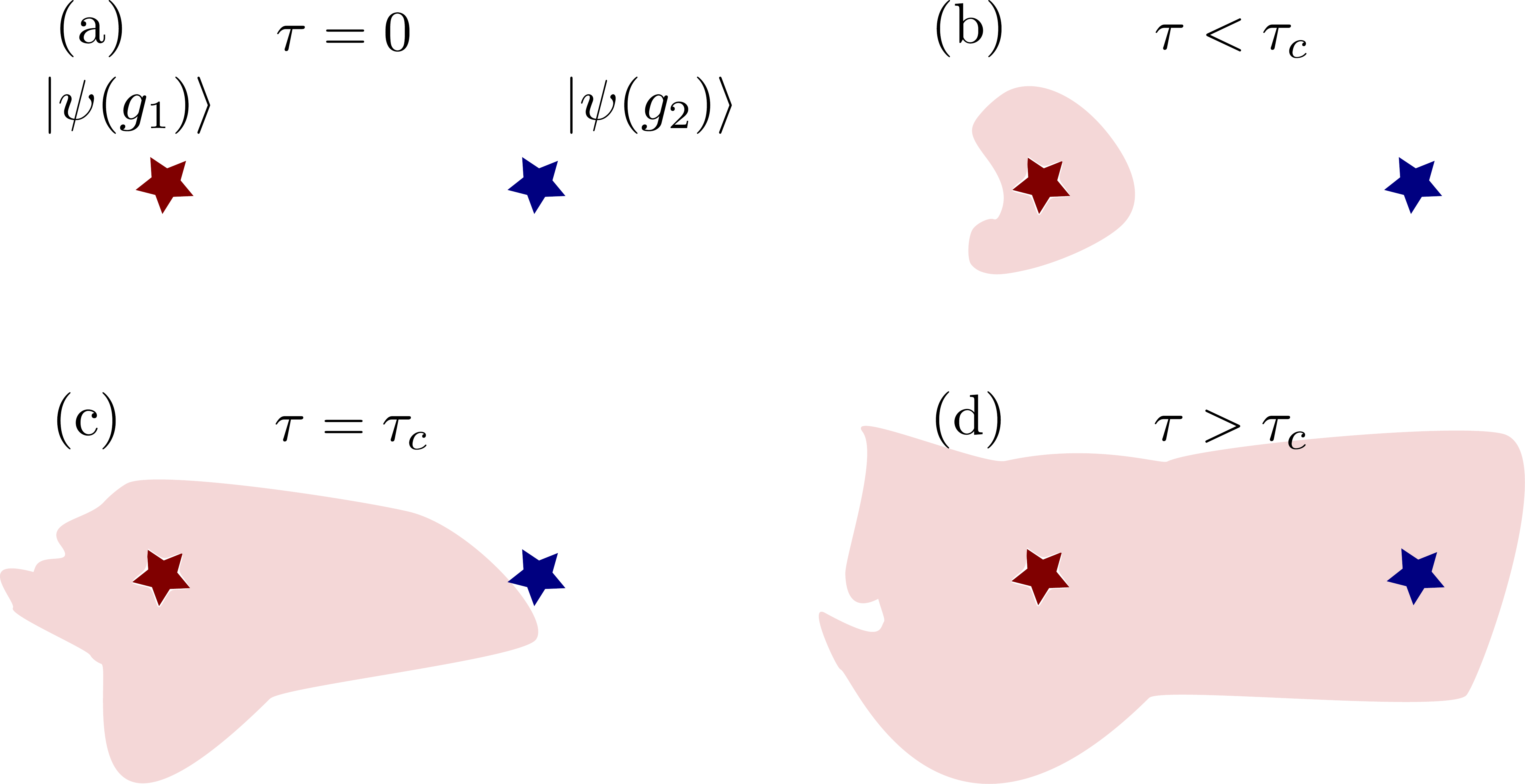,width=8cm}}
\vspace*{8pt}
\caption{(a) The initial state and the target state. We can not change the state if $\tau=0$ so the set of all reachable states is just the initial state. (b) The set of reachable states grows continuously in the multidimensional space of wave functions. (c) At a critical time $\tau_c$, the final state becomes reachable. (d) For $\tau>\tau_c$, we expect many protocols to prepare $|\psi(g_2)\rangle$.\label{f3}}
\end{figure}

As a simple application, in this section, we review a dynamical phase transition found in the preparation of the ground state of a Luttinger liquid.\cite{Rahmani2011}  Before discussing this specific system, let us argue why we expect such a dynamical transition generically. Let us assume we are in the ground state $|\psi(g_1)\rangle$ of a Hamiltonian $H(g_1)$. We change the parameter $g(t)$ for a time $\tau$. The goal is to reach the ground state of $H(g_2)$. As discussed earlier, the wave function can be labeled by a set of numbers (such as amplitudes in a fixed basis), which define a multidimensional space. The initial wave function corresponds to a point in this space. If $\tau=0$, it is impossible to change the wave function and the reachable set of final wave functions only includes $|\psi(g_1)\rangle$ as shown in Fig.~\ref{f3}(a). As we increase $\tau$, the set of all final wave functions, which can be reached by all permissible protocols over a time $\tau$, defines a region of the wave-function space. We expect this region to be continuous, include $|\psi(g_1)\rangle$ and grow as function of $\tau$. Initially, the point $|\psi(g_1)\rangle$ lies outside this region. [Fig.~\ref{f3}(b)] At some critical time $\tau_c$, this growing region (the reachable set of final wave functions) hits the target wave function $|\psi(g_2)\rangle$ as shown in Fig.~\ref{f3}(c). In the optimal control framework, we fix $\tau$, and try to find protocols that bring the wave function as close as possible to  the target state $|\psi(g_2)\rangle$ by maximizing a figure of merit such as the overlap of the final state with $|\psi(g_2)\rangle$ (which will have a maximum of identity when the state is reached exactly). Thinking of a corresponding cost function, which is positive for  $|\psi(g_2)\rangle\neq |\psi(\tau)\rangle$ and vanishes for $|\psi(g_2)\rangle= |\psi(\tau)\rangle$, as an order parameter, the critical time $\tau_c$ corresponds to a dynamical phase transition, at which the order parameter vanishes. For $\tau>\tau_c$, we expect many different protocols to prepare $|\psi(g_2)\rangle$ exactly.

Using Eq. (\ref{eq:ricc}), the dynamics of a Luttinger liquid was studied for an arbitrary protocol, and the simulated annealing method was used to find optimal protocols, which, for a given initial state, maximize the overlap of the final state with a given final target state.\cite{Rahmani2011} The transition explained above was found in the numerical simulations for a $\tau_c \propto L$, where $L$ is the length of system. After such critical time, the numerically obtained cost function plunges down by several orders of magnitude indicating that the exact preparation of the target state is possible for such times. It is important to notice that such dramatic change is only observed in the cost function obtained by optimal protocols, and simple linear or power-law protocols do not capture this change of behavior. An interesting question is what sets the scaling of $\tau_c$ with the system size. As the evolution is optimal, we expect $\tau_c$ to be set by the quantum speed limit, which in general is related to the spread  (uncertainty) of energy in the quantum wave function.\cite{Battacharyya1983,Anandan1990,Pfeifer1993,Giovannetti2003,Carlini2006,Rezakhani2009,Caneva2009}

In case of the Luttinger liquid, the energy gap to excitations scales as $\Delta E\sim 1/L$, which indicates $\tau_c \propto 1/\Delta E$. Based on similar criteria for adiabatic evolution (even though the optimal protocols are highly nonadiabatic),\cite{Schaller2006} it is tempting to conjecture that $\tau_c \propto 1/\Delta E$ may be a generic result. In fact by considering systems with a quantum critical point and one tunable parameter, this result has been shown to be more general.\cite{Caneva2011} However, the problem needs to  be defined very precisely in terms of the changes we are allowed to make to the Hamiltonian. In the Luttinger-liquid case, one parameter (nearest-neighbor interaction strength) was allowed to change in a fixed range. It is known, from the theory of transitionless quantum driving~\cite{Berry2009}, that that $\tau_c$ can be made ${\cal O}(1)$ if we are allowed to add long-range interactions to the Hamiltonian. Even adding a larger number of local interactions can in some cases change the scaling of $\tau_c$. For example  in a transverse field Ising model with ferromagnetic interactions for $z$ components of the spins and a transverse field in the $x$ direction, we would need a time of order  $1/\Delta E$ to go across the quantum critical point by tuning the transverse field.\cite{Caneva2011} However, we can avoid going directly across the critical point by adding a secondary transverse field in the $y$ direction, which can simply rotate a product state with all spins in the $x$ direction (paramagnetic side of the critical point) to a product state with all spins in the $z$ direction (ferromagnetic side of the critical point) in time of order  ${\cal O}(1)$.

\subsection{Cooling through optimal control}

The optimal-control approach outlined above can also be applied to mixed states described by density matrices.\cite{Rahmani2013} While creating model Hamiltonians such as the Hubbard model has been successful, a long-standing challenge in the quantum simulation of condensed matter models with cold atoms is cooling the atomic systems to extremely low temperatures so that the ground state properties can be observed (e.g. antiferromagnetism or the putative $d$-wave superconductivity in the fermionic Hubbard model). As optimal control is a powerful tool for transforming the quantum states into more desirable states, the question arises as to whether such schemes can be used for cooling. As expected, the second law of thermodynamics places important restrictions on any cooling scheme.

In fact, if we have a thermal state $\rho=e^{-\beta H_0}/Z$, where $\beta$ is the inverse temperature, we cannot have a cyclic process $H(t)$ with $H(0)=H(\tau)=H_0$ such that the corresponding unitary evolution reduces the expectation value of energy with respect to the Hamiltonian $H_0$.\cite{Rahmani2013,Thirring2002} In other words, one can show that 
\begin{equation}\label{eq:ineq}
 {\rm tr}\left( H_0 \rho_0 \right) \leqslant {\rm tr}\left(H_0 U^\dagger\rho_0 U\right),
\end{equation}
for any unitary operator $U$. As the unitary time evolution is reversible, the irreversibility encoded in the thermodynamic inequality above stems solely from the properties of the initial density matrix $\rho_0$, which is assumed thermal. The density matrices satisfying the inequality above are known as passive density matrices.\cite{Thirring2002}

 A sufficient condition for passivity is being (i) diagonal in the basis of the eigenstates of $H_0$ and (ii) $\rho_{ii}\geqslant\rho_{jj}$ for $E_{i}\leqslant E_{j}$, where $E_{i}$ is an eigenvalue of $H_0$. For such density matrices, the initial average energy [left-hand side of Eq.~(\ref{eq:ineq})] is given by $\sum_i E_i\rho_{ii}$. After the unitary evolution with an evolution operator $U$ (with matrix elements $U_{ij}$ in the same energy basis), the final energy expectation value [right-hand side of Eq.~(\ref{eq:ineq})] is given by $\sum_{ij} E_i W_{ij}\rho_{jj}$, where $W_{ij}\equiv U^{\dagger}_{ij}U_{ji}$. We now invoke an important theorem in discrete mathematics known as the von Neumann-Birkhoff theorem\cite{Graham1995}, which states that any square matrix $A$ of nonnegative real numbers with the following property
\begin{equation}
\sum_i A_{ij}=\sum_j A_{ij}=1,
\end{equation}
i.e., all rows and columns adding up to identity, can be written as a sum $A=\sum_k c_k P_k$ of permutation matrices $P_k$ such that with $\sum_k c_k=1$ (the coefficients $c_k$ are positive).

The matrix $W$ clearly satisfies the above condition (such matrices are referred to as doubly stochastic) as its elements are transition probabilities. Now according to von Neumann-Birkhoff theorem, we have $W=\sum_k c_k P_k$. Therefore, the right-hand side of  Eq.~(\ref{eq:ineq}) can be written as $\sum_{ij} E_i W_{ij}\rho_{jj}=\sum_{ijk} c_kE_i P^k_{ij}\rho_{jj} $.
Condition (ii) of passivity implies that, for any permutation of the weights $\rho_{ii}$, $\sum_{ij} E_i P^k_{ij}\rho_{ii} \geqslant \sum_{i} H_i \rho_{ii}$. Since $\sum_k c_k=1$, we immediately obtain our inequality.

\begin{figure}[th]
\centerline{\psfig{file=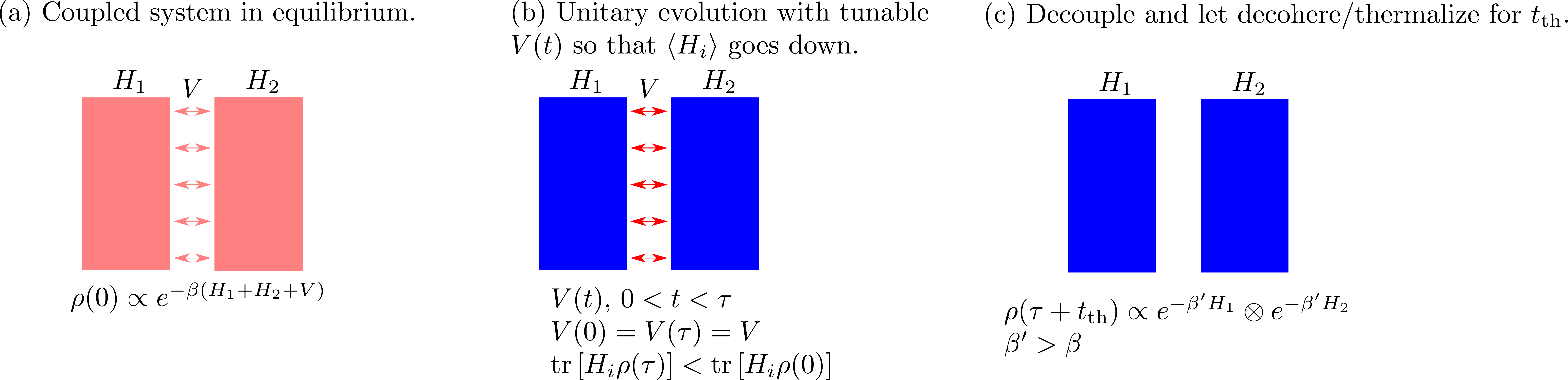,width=12cm}}
\vspace*{8pt}
\caption{(a) Strongly couple the two systems and cool the coupled system with other methods so it equilibrates at the lowest possible temperature. (b) Perform unitary evolution on the whole system by making the coupling time-dependent so that the expectation value of $H_1$ decreases (by optimal control). The time scales must be short enough that the system remains quantum coherent. (c) Suddenly decouple the systems (set the coupling to zero) and let it equilibrate.\label{f4}}
\end{figure}
It appears that we have a no-go theorem, which makes it impossible to use unitary evolution for cooling. However, we can get around this by enlarging the system so that it is a subsystem of a larger systems. This can be achieved, e.g., by creating two copies of the system we want to cool, and coupling them via certain couplings $H=H_1+H_2+V$, where $H_{1,2}$ represents the Hamiltonian of each copy and $V$ is the coupling Hamiltonian. Using this trick, we effectively eliminate the restrictions of cyclicity as the Hamiltonian of each copy can be cyclic (even remain constant throughout the evolution). We can only make use of variations of the coupling $V$ to generate the unitary evolution. At the end of the process, we want to have just the system $H_1$ with a reduced amount of excess energy so we would like to decouple the systems at the end (very fast so the density matrix remains the same and no energy is spilled into the the subsystem $H_1$).

The scheme utilizes the finite coherence times of quantum systems. The unitary evolution is used over a time $\tau$, where the entire system is quantum coherent and the subsystems remain entangled even when they are decoupled (we may need to set $V(t)=0$ sometimes during the evolution $0<t<\tau$). Once the unitary transformation is done (if the transformation up to this point is cyclic, we have $V(\tau)=V$), the expectation value $\langle H_1+H_2+V\rangle$ can not have decreased. However, it is possible that $H_i$ expectation values have decreased at the expense of an increased $\langle V\rangle$ (the expectation values are defined as $\langle A\rangle\equiv{\rm tr}\left[A \rho(\tau)\right]$).

When the coupling $V$ is set to zero for a final time, the scheme relies on the thermalization of system $H_1$: We just managed to reduce the expectation value of energy of $H_1$ below what it would be at inverse temperature $\beta$. The low energy state prepared right after the unitary process (at time $\tau$) is not, however, in equilibrium. If we wait some additional time $t_{\rm th}$ though, the system should redistribute its (lowered) excess energy and equilibrate to a lower temperature. In addition to the final thermalization, the reduction of energy through unitary evolution relies on the free flow of energy in the system. We thus expect such cooling scheme to fail for systems exhibiting many-body localization, and apply to all other systems.

This method has been (theoretically) applied to a specific system comprised of two elongated quasicondensates (fluctuating one-dimensional condensates).\cite{Rahmani2013} It was found that significant cooling can be achieved. Remarkably, the protocols obtained for this cooling scheme by an unbiased simulated annealing calculation have a bang-bang structure as expected from the Pontryagin's maximum principle.\cite{Rahmani2013} According to this theorem, whenever we have a linear control $\alpha_{\rm min}<\alpha(t)<\alpha_{\rm max}$, (in case of quantum evolution, the Hamiltonian is linear in the coupling constant $\alpha(t)$ that we tune to generate the optimal evolution), the optimal  $\alpha(t)$ can only take the limiting values $\alpha_{\rm min}$ and $\alpha_{\rm max}$, and the coupling constant should jump between these values though a sequence of sudden quenches (this is the generic behavior for linear control but the theorem does allow for exceptions).

\section{Summary}\label{sec:conc}
In summary, we reviewed several ideas and techniques, which have emerged recently in nonequilibrium quantum dynamics of thermally isolated many-body systems. The key ingredient in problems surveyed in this article is the fact that we need to understand an ensemble of quantum Hamiltonians as opposed to time evolution with a single time-dependent Hamiltonian. We divided these problems into three categories: (i) spatially disordered systems characterized by many-body localization, where the slow growth of entanglement entropy has been identified as a useful numerical signature of the many-body localization transition. (ii)  noise-driven quantum systems, which arise naturally in optical lattices. As the lattices are formed by lasers, which generally have time-dependent fluctuations, the  Schr\"odinger equation governing the dynamics is stochastic. We discussed a way to characterize the fluctuations of absorbed energy in such systems, while distinguishing the contributions of quantum and stochastic fluctuations. We also described two approaches to the problem based on the Fokker-Planck and master equations, which can be used in analyzing the effects of noise on energy and other observables such as the correlation functions. As a concrete example, we discussed the application of the Fokker-Planck approach to a noisy Luttinger liquid in some detail. Finally we turned to a setup (iii) with a rather different philosophy, namely, the selection of optimal protocols out of an ensemble of permissible ones. We argued that very direct simulated-annealing calculations can indeed yield such optimal protocols. One advantage of the simulated annealing is that in can be incorporated into the experiment.

The three directions above are by no means the only problems arising in ensemble quantum evolution. The field is largely unexplored, and appears to contain rich physics both from practical and fundamental points of view. In addition to exploring the three questions above for different systems, combinations of the above questions can be raised: (a) what is the fate of many-body localized state when driven by a noisy Hamiltonian? Constant driving of the system appears to lead to more heating that a sudden quench. Can it overcome the localization transition? (b) Can one perform optimal control in the presence of strong disorder (localization)?
(c) What about optimal control in the presence of noise? How robust are optimal control protocols to noise? Can we design optimal protocols, which effectively cancel the unwanted heating effects of noisy optical lattices? And finally (d), putting all these together, how can we do optimal control in the presence of both spatial and temporal disorder? The progress over the last few years suggests that applying the methods of classical statistical mechanics to problems involving ensemble quantum evolution of thermally isolated systems may give rise to a deeper understanding of fundamentals of quantum statistical mechanics and pave the way for further experimental developments in quantum simulations.

\section*{Acknowledgments}

I am grateful to Claudio Chamon, Luca D'Alessio, Eugene Demler, and Takuya Kitagawa for collaboration in research reviewed in this article. I thank Ehud Altman for his careful reading of the manuscript and several helpful suggestions, and David Huse for important comments and discussions. This work was supported by the U.S. Department of Energy under LANL/LDRD program.


\begin{thebibliography}{0}


\bibitem{Greiner2002}
M. Greiner \textit{et al}., Nature (London) \textbf{419}, 51 (2002).

\bibitem{Tuchman2006}
A. K. Tuchman \textit{et al}., Phys. Rev. A \textbf{74}, 051601 (2006).
 
\bibitem{Kinoshita2006}
T. Kinoshita, T. Wenger, and D. S. Weiss, Nature (London) \textbf{440}, 900 (2006).

\bibitem{Sadler2006}
L. E. Sadler, J. M. Higbie, S. R. Leslie, M. Vengalattore and D. M. Stamper-Kurn,
Nature \textbf{443}, 312 (2006).

\bibitem{Hofferberth2007}
 S. Hofferberth \textit{et al}., Nature (London) \textbf{449}, 324 (2007).




\bibitem{Lewenstein2007}
M. Lewenstein, A. Sanpera, V. Ahufinger, B. Damski, A. Sen, and U. Sen, Adv. Phys. \textbf{56}, 243 (2007).

\bibitem{Bloch2008}
I. Bloch, J. Dalibard and W. Zwerger, Rev. Mod. Phys. \textbf{80}, 885 (2008).

\bibitem{Polkovnikov2011}
A. Polkovnikov, K. Sengupta, A. Silva, and M. Vengalattore, Rev. Mod. Phys. \textbf{83}, 863 (2011).

\bibitem{Znidaric2008}
M. \u{Z}nidari\u{c}, T. Prosen, and P. Prelov\u{s}ek, Phys. Rev. B \textbf{77}, 064426 (2008).

\bibitem{Chiara2006}
G. D. Chiara, S. Montangero, P. Calabrese, and R. Fazio, J. Stat. Mech., P03001 (2006).


\bibitem{Bardarson2012}
J. H. Bardarson, F. Pollmann, and J. E. Moore, Phys. Rev. Lett. \textbf{109}, 017202 (2012).

\bibitem{Khatami2012}
E. Khatami, M. Rigol, A. Rela\~no, and A. M. Garc\'ia-Garc\'ia, Phys. Rev. E \textbf{85}, 050102(R) (2012).

\bibitem{Vosk2013}
R. Vosk and E. Altman, Phys. Rev. Lett. \textbf{110}, 067204 (2013).


\bibitem{Serbyn2013}
M. Serbyn, Z. Papi\'c, and D. A. Abanin, Phys. Rev. Lett. \textbf{110}, 260601 (2013).

\bibitem{Huse2013}
D. A. Huse and V. Oganesyan, arXiv:1305.4915.

\bibitem{Basko2006}
D. M. Basko, I. L. Aleiner, and B. L. Altshuler, Annals of Physics \textbf{321}, 1126 (2006).

\bibitem{Basko2007}
D. M. Basko, I. L. Aleiner, and B. L. Altshuler, Phys. Rev. B \textbf{76}, 052203 (2007)


\bibitem{Oganesyan2007}
V. Oganesyan and D. A. Huse, Phys. Rev. B \textbf{75}, 155111 (2007).

\bibitem{Pal2010}
A. Pal and D. A. Huse, Phys. Rev. B \textbf{82}, 174411 (2010).



\bibitem{Dalessio2013}
L. D'Alessio and A. Rahmani, Phys. Rev. B \textbf{87}, 174301 (2013).


\bibitem{Pichler2012}
H. Pichler, J. Schachenmayer, J. Simon, P. Zoller, and A. J. Daley, Phys. Rev. A \textbf{86}, 051605(R) (2012).

\bibitem{Pichler2013}
H. Pichler, J. Schachenmayer, A. J. Daley, and P. Zoller, Phys. Rev. A \textbf{87}, 033606 (2013).


\bibitem{DallaTorre2010}
10E. G. Dalla Torre, E. Demler, T. Giamarchi, and E. Altman, Nature
Phys.\textbf{ 6}, 806 (2010).

\bibitem{DallaTorre2012}
E. G. Dalla Torre, E. Demler, T. Giamarchi, and E. Altman, Phys. Rev. B \textbf{85}, 184302 (2012).


\bibitem{Marino2012}
J. Marino and A. Silva,  Phys. Rev. B \textbf{86}, 060408(R) (2012).



\bibitem{Doria2011}
P. Doria, T. Calarco, and S. Montangero, Phys. Rev. Lett. \textbf{106}, 190501 (2011).

\bibitem{Rahmani2011}
A. Rahmani and C. Chamon, Phys. Rev. Lett. \textbf{107}, 016402 (2011).


\bibitem{Montangero2007}
S. Montangero, T. Calarco, and R. Fazio, Phys. Rev. Lett. \textbf{99}, 170501 (2007). 


\bibitem{Brif2010}
C. Brif, R. Chakrabarti, and H. Rabitz, New J. Phys. \textbf{12}, 075008 (2010). 

\bibitem{Peirce1988}
A. P. Peirce, M. A. Dahleh, and H. Rabitz, Phys. Rev. A \textbf{37}, 4950 (1988). 

\bibitem{Calarco2004}
T. Calarco, U. Dorner, P. Julienne, C. J. Williams, and P. Zoller, Phys. Rev. A \textbf{70}, 012306 (2004). 

\bibitem{Hohenester2007}
U. Hohenester, P. K. Rekdal, A. Borzi, and J. Schmiedmayer, Phys. Rev. A \textbf{75}, 023602 (2007). 

\bibitem{Grond2009}
 J. Grond, J. Schmiedmayer, and U. Hohenester, Phys. Rev. A \textbf{79}, 021603 (2009). 

\bibitem{Rahmani2013}
A. Rahmani, T. Kitagawa, E. Demler, and C. Chamon, Phys. Rev. A \textbf{87}, 043607 (2013).

\bibitem{Torrontegui2013}
E. Torrontegui \textbf{et al}, Adv. At. Mol. Opt. Phys. \textbf{62}, 117 (2013).


\bibitem{Chen2010}
X. Chen, A. Ruschhaupt, S. Schmidt, A. del Campo, D. Guéry-Odelin, and J. G. Muga, Phys. Rev. Lett. \textbf{104}, 063002 (2010).

\bibitem{Masuda2010}
S. Masuda and K. Nakamura, Proc. R. Soc. A \textbf{466}, 1135 (2010).

\bibitem{Berry2009}
M. V. Berry, J. Phys. A \textbf{42}, 365303 (2009).

\bibitem{Rigol2008}
M. Rigol, V. Dunjko and M. Olshanii, Nature \textbf{452}, 854 (2008).


\bibitem{Anderson1958}
P. W. Anderson, Phys. Rev. \textbf{109}, 1492 (1958).



\bibitem{Calabrese2005}
P. Calabrese and J. Cardy, J. Stat. Mech. (2005) P04010.


\bibitem{Fagotti2008}
M. Fagotti and P. Calabrese, Phys. Rev. A \textbf{78}, 010306(R) (2008).

\bibitem{Sengupta2009}
K. Sengupta and D. Sen, Phys. Rev. A \textbf{80}, 032304 (2009). 

\bibitem{Rahmani2010}
A. Rahmani, and C. Chamon, Phys. Rev. B \textbf{82}, 134303 (2010). 






\bibitem{Kim2013}
H. Kim and D. A. Huse, Phys. Rev. Lett. \textbf{111}, 127205 (2013).

\bibitem{Bauer2013}
B. Bauer and C. Nayak, arXiv:1306.5753.


\bibitem{Deutsch1991}
J. M. Deutsch, Phys. Rev. A \textbf{43}, 2046 (1991).


\bibitem{Srednicki1994}
M. Srednicki, Phys. Rev. E \textbf{50}, 888 (1994).



\bibitem{Dasgupta1980}
C. Dasgupta and S.-K. Ma, Phys. Rev. B \textbf{22}, 1305 (1980).

\bibitem{Bhatt1982}
R. N. Bhatt and P. A. Lee, Phys. Rev. Lett. \textbf{48}, 344 (1982).

\bibitem{Fisher1994}
D. S. Fisher, Phys. Rev. B \textbf{50}, 3799 (1994).

\bibitem{Refael2004}
G. Refael and J. E. Moore, Phys. Rev. Lett. \textbf{93}, 260602 (2004).



\bibitem{Vosk2013b}
R. Vosk and E. Altman, arXiv:1307.3256.

\bibitem{Choi2012}
S. Choi, R. Onofrio, and B. Sundaram, Phys. Rev. A \textbf{86}, 043436 (2012).

\bibitem{Li2013}
J. Li \textit{et al}, Nat. Commun. 4, 1420 (2013).
\bibitem{Bunin2011}
G. Bunin, L. D'Alessio, Y. Kafri, and A. Polkovnikov, Nat. Phys. \textbf{7}, 913 (2011). 

\bibitem{Molmer1993}
K. M\"olmer, Y. Castin, and J. Dalibard, J. Opt. Soc. Am. B \textbf{10}, 524 (1993). 

\bibitem{Polkovnikov2008}
A. Polkovnikov and V. Gritsev, Nature Phys. \textbf{4}, 477 (2008).

\bibitem{Risken1989} H. Risken, \textit{The Fokker-Planck Equation: Methods of
Solutions and Applications Springer} (Springer, New York, 1989). 

\bibitem{Schaller2006}
G. Schaller, S. Mostame, and R. Schützhold, Phys. Rev. A \textbf{73}, 062307 (2006). 

\bibitem{Cirac2012}
J. I. Cirac and P. Zoller, Nature Phys. \textbf{8}, 264 (2012); I. Bloch, J. Dalibard and S. Nascimbne, ibid. \textbf{8}, 267 (2012); R. Blatt and C. F. Roos, ibid. \textbf{8}, 277 (2012).


\bibitem{Battacharyya1983}
K. Battacharyya, J. Phys. A \textbf{16}, 2993 (1983).

\bibitem{Anandan1990}
J. Anandan and Y. Aharonov, Phys. Rev. Lett. \textbf{65}, 1697 (1990).

\bibitem{Pfeifer1993}
P. Pfeifer, Phys. Rev. Lett. \textbf{70}, 3365 (1993).

\bibitem{Giovannetti2003}
V. Giovannetti, S. Lloyd, and L. Maccone, Phys. Rev. A \textbf{67}, 052109 (2003).



\bibitem{Carlini2006}
A. Carlini, A. Hosoya, T. Koike, and Y. Okudaira, Phys. Rev. Lett. \textbf{96}, 060503 (2006).


\bibitem{Rezakhani2009}
A. T. Rezakhani, W.-J. Kuo, A. Hamma, D. A. Lidar, and P. Zanardi, Phys. Rev. Lett. \textbf{103}, 080502 (2009).


\bibitem{Caneva2009}
T. Caneva, M. Murphy, T. Calarco, R. Fazio, S. Montangero, V. Giovannetti, and G. E. Santoro, Phys. Rev. Lett. \textbf{103}, 240501 (2009).

\bibitem{Caneva2011}
T. Caneva, T. Calarco, R. Fazio, G. E. Santoro, and S. Montangero Phys. Rev. A \textbf{84}, 012312 (2011). 

\bibitem{Thirring2002}
W. Thirring, \textit{Quantum Mathematical Physics: Atoms, Molecules and Large Systems} (Springer, Berlin, 2002). 

\bibitem{Graham1995}
R. Graham, M. Gr\"otschel, and L. Lav\'asz, \textit{Handbook of Combinatorics }(Elsevier, Amsterdam, 1995). 


\end{thebibliography}
\end{document}